\documentclass[twocolumn,nofootinbib,
preprintnumbers,superscriptaddress]{revtex4-1}

\usepackage{amsmath,amssymb,amsfonts,amsthm,graphicx, epsf,dsfont,dcolumn,yfonts}
\usepackage{latexsym}
\usepackage{bm}
\usepackage{braket}
\newcommand{\mb}{\mathbf}
\usepackage[colorlinks=true, pdfstartview=FitV, linkcolor=black, citecolor=black, urlcolor=black]{hyperref}
\usepackage{slashed} 
\usepackage{color}
\usepackage{slashed} 
\newcommand{\be}{\begin{equation}}
\newcommand{\ee}{\end{equation}}
\newcommand{\bea}{\begin{eqnarray}}
\newcommand{\eea}{\end{eqnarray}}
\newcommand{\ba}{\begin{align}}
\newcommand{\ea}{\end{align}}
\def\ie{{\it i.e.~}}

\newcommand{\bwt}{\begin{widetext}}
\newcommand{\ewt}{\end{widetext}}

\def\T{{\rm T}}

\def\det{{\rm{det}}}

\def\nf{{n_{\rm F}}}

\newcommand{\Oi}{\mathcal{O}}

\begin{document}
\title{Dyonic Stars for Holography}
\author{Daniel Carney}
\affiliation{Theory Group and Texas Cosmology Center \\ University of Texas at Austin, Austin TX 78712, USA}
\affiliation{Department of Physics and Astronomy and Pacific Institute of Theoretical Physics \\ University of British Columbia, Vancouver BC V6T 1Z1, Canada}
\author{Mohammad Edalati}
\affiliation{Theory Group and Texas Cosmology Center \\ University of Texas at Austin, Austin TX 78712, USA}

\begin{abstract}
We construct back-reacted asymptotically AdS$_4$ backgrounds with both electric and magnetic charge, at finite temperature and chemical potential. The system consists of a condensed bulk Dirac field hovering over a charged black brane. We give a detailed microscopic treatment of the bulk fermions, from which we derive the usual fluid approximations describing the condensate. In the context of holography, such a gravitational solution is dual to a (2+1)-dimensional CFT with a $U(1)$ current, at finite density and strong coupling, subjected to a transverse magnetic field. 

\end{abstract}

\maketitle

\tableofcontents

\section{Introduction}
Consider a high density system of fermions in $2+1$ dimensions with a global $U(1)$ symmetry, at strong coupling and subject to a transverse magnetic field $B$. Assuming such a system could be given a holographic description, what is the dual gravitational solution? This paper proposes an answer to this question. 

To model such a system holographically, one is searching for a gravity solution in asymptotically AdS$_4$ spacetime with a $U(1)$ gauge field and a weakly coupled bulk fermion satisfying suitable boundary conditions. Treating the fermions in the probe limit corresponds to a boundary system in which the thermodynamics is controlled entirely by external sources. In order to make connections with realistic systems and to learn about the microscopics of the charges making up the state, one needs to include the stress-energy of the bulk fermions in the gravity solution. 

In this paper, we construct bulk solutions with these properties and with the fermions and gauge field contributing to the stress-energy at the same order, at arbitrary magnetic field strength. The solutions consist of a large number of bulk fermionic states localized to a finite region above a black brane carrying both electric and magnetic charge. We call these solutions dyon stars. 

Fermionic systems at finite density and magnetic field are critically important in condensed matter physics. The best-understood examples are at weak coupling, where the fermion spectrum consists of quantized cyclotron orbits. This leads to a number of fascinating phenomenon, especially the de Haas-van Alphen oscillations in thermodynamic quantities in $1/B$. Moreover, the ground state of the system is degenerate even at zero temperature. However, it is far from obvious what parts of this picture carry over to the strong-coupling regime. One can study the problem using the AdS/CFT correspondance, which requires solutions of the type presented in this work.

Early work focused on probe fermions in electrically charged black brane backgrounds \cite{Hartnoll:2007ih}. More recently, probes in magnetically charged black brane backgrounds \cite{Hartnoll:2007ai,Allais:2012ye}, including at parametrically large $B/\mu^2$ \cite{Blake:2012tp}, have been studied. Beyond the probe approximation, there are now solutions including backreaction of charged \cite{Hartnoll:2009ns,Hartnoll:2010gu,Hartnoll:2010ik,Allais:2013lha} and neutral \cite{deBoer:2009wk} fermions, without bulk magnetic fields. Here we fill the obvious gap by constructing the solutions at arbitrary $B/\mu^2 \lesssim 1$, including backreaction from charged fermions. We will work at small but finite temperature $0 < T/\mu \ll B/\mu^2$.

The dyon star solutions exhibit a rich and interesting phenomenology. They display oscillations in their density profiles along the holographic direction as a result of the bulk dHvA effect. Like the purely electron stars \cite{Sachdev:2011ze,Hartnoll:2010ik}, the dyon stars are thermodynamically favored over pure black brane backgrounds at temperatures below a critical temperature $T_c = T_c(B,\mu)$, and the respective free energies of the solutions suggest a third-order phase transition from the brane solutions to the stars.

Yesterday \cite{Puletti:2015gwa} appeared, which overlaps strongly with the results presented here; the main difference is that we have given a detailed statistical-mechanical treatment of the bulk fluid. While this work was in progress the paper \cite{Burikham:2012kn} appeared, which covers the same parameter space but uses a very different set of approximations, leading the authors to qualitatively different solutions than those presented here.

\section{Strategy and bulk equations}
We are looking for bulk solutions with AdS asymptotics, significant fermionic backreaction, and a gauge field with both radial magnetic field and chemical potential extending out to the boundary. At either zero or finite temperature, the easiest way to engineer the gauge field is to assume the existence of a black brane with both electric and magnetic charges deep in the bulk. Then we can start populating fermionic modes in this background by turning on a boundary chemical potential, searching for a self-consistent solution in which these fermions backreact on both the geometry and gauge field.

Just as in an astrophysical fermionic star, the gravitational forces acting to condense the fermions must fight against degeneracy pressure, which can ultimately stabilize the fermions against collapse. This requires a high density of fermions, which in turn means that one can view them as an effective fluid obeying some particular equation of state. Then the gravitational dynamics in the region populated by the fermions can be described in the Tolman-Oppenheimer-Volkoff approximation \cite{tolman,oppenheimer}. In turn, the equations of motion for the fermionic modes can be solved in a WKB approximation along the radial direction \cite{Ruffini:1969qy}. In these approximations, the coupled Einstein-Maxwell-Dirac system can be reduced to a set of coupled ODEs along the holographic direction.

Unlike a typical astrophysical star, here we have begun with a charged black brane sitting deep in the AdS bulk. We are anticipating that the fermions can condense above this horizon. If the black brane is only electrically charged, it is known that at $T>0$ it is thermodynamically favored for the fermions to condense above the horizon, and at $T=0$ the condensed fermions ``fall into'' the black brane and one obtains a Lifshitz-type geometry in which the horizon has dissolved into an ``electron star''.\cite{Hartnoll:2010gu,Allais:2013lha} However, if the horizon carries magnetic charge, this is not possible: because the black brane acts as a magnetic monopole, the horizon cannot go away without providing some other source for the magnetic field. In \cite{Albash:2012ht} this was achieved by adding a dilaton, but for purposes of minimality here we will only consider the bulk fermion and gauge field, thus all of our solutions have a black brane horizon in the deep IR.

In the bulk, we consider a $(3+1)$-dimensional system consisting of the metric $\underline{g}_{\mu\nu}$, an Abelian gauge field $\underline{A}_{\mu}$, and a charged (Dirac) fermion $\underline{\psi}$; the underline differentiates these fields from their rescaled partners to be defined shortly. The Einstein-Maxwell-Dirac system is governed by
\begin{align}\label{BulkLagrangian}
S = \int d^4\underline{x}\sqrt{-\underline{g}}\Big\{&\frac{1}{2\kappa^2}\left(\underline{R} - 2 \Lambda\right) -\frac{1}{4q^2}  \underline{F}^2\nonumber\\
&- i \underline{\overline{\psi}} (\underline{\slashed{D}}-\underline{m}) \underline{\psi}\Big\},
\end{align}
where $\kappa^2=8\pi G_{\rm N}$, $\underline{R}$ is the Ricci scalar, $\Lambda = -3/L^2$ is the cosmological constant, $\underline{F}=\underline{d}\underline{A}$ is the $U(1)$ field strength, $q$ and $\underline{m}$ are, respectively, the charge and the mass of bulk spinor $\underline{\psi}$. Our conventions for the spin geometry, covariant derivatives, etc. are spelled out in appendix \ref{spingeo}. Throughout this paper, Greek indices $\mu,\nu$ refer to bulk spacetime coordinates while hatted indices $\hat{a},\hat{b}$ refer to bulk tangent frame indices. As explained in the appendix, we will repeatedly use veilbeins $e_{\hat{a}}^{\mu}$ to write quantities evaluated in the tangent frame; for example the local momentum is $p_{\hat{a}} = e^{\mu}_{\hat{a}} p_{\mu}$.

We are interested in the regime in which the gauge and fermionic terms in this action backreact significantly on the gravitational solution. To do this in the large $N$ limit, it is convenient to rescale various quantities. There are two dimensionless couplings in this system, which we denote\footnote{The parameter $\beta^2$ here is morally equivalent to the phenomenological parameter $\hat{\beta}$ used in \cite{Hartnoll:2007ai,Hartnoll:2010gu,Hartnoll:2010ik}; since we are treating the fermions microscopically we can give a precise formula for the coefficient.} 
\be
\gamma = \frac{qL}{\kappa}, \ \ \beta = q \gamma.
\ee
Define dimensionless coordinates $x^{\mu} = \underline{x}^{\mu}/L$. Now we define rescaled, dimensionless fields and derivatives
\begin{align}
\begin{split}
R & = L^2 \underline{R}, \ \ \psi = \left( \frac{L}{\gamma} \right)^{3/2} \underline{\psi}, \ \ m = \frac{L}{\gamma} \underline{m} \\
A_{\mu} & = \frac{L}{\gamma} \underline{A}_{\mu}, \ \ D_{\mu} = \frac{L}{\gamma} \underline{D}_{\mu} = \frac{1}{\gamma} \partial_{\mu} - \frac{i L}{4\gamma} \underline{\omega}_{\mu,\hat{a}\hat{b}} \Gamma^{\hat{a}\hat{b}} - i A_{\mu}.
\end{split}
\end{align}
In terms of these rescaled fields and $F=dA$ we have the action expressed in terms of our dimensionless variables:
\begin{align}\label{BulkLagrangianDimLessFinal}
S = \frac{L^2}{\kappa^2}\int d^4 x\sqrt{-g}\Big\{ \frac{1}{2}\left(R +6\right) -\frac{1}{4} {F}^2- i \beta^2 \overline{\psi} (\slashed{D}-m) \psi \Big\}. 
\end{align}
In writing the action this way, we have scaled out the large $N \sim L/\kappa$ behavior, and we will work in the regime where all the terms in the action contribute at the same order. We will often set $L=1$ without loss of generality. 

We would like to work in a regime with a large density of occupied fermion states. This constrains our parameters $\gamma,\beta$. First note that in holography, one requires $L/\kappa = \gamma^2/\beta \gg 1$. Moreover, we will see that the WKB approximation for the fermions, or equivalently the locally flat fluid approximation, requires $\gamma \gg 1$.\footnote{In this sense, here the TOV approximation is equivalent to the  Thomas-Fermi \cite{thomas,fermi} approximation.} To obtain an order-one backreaction due to the fluid, we want $\beta^2 \sim \Oi(1)$. Putting these constraints together means that we are working with $q^2 \ll 1$.

The equations of motion coming from \eqref{BulkLagrangianDimLessFinal} read
\begin{align}
\begin{split}
G_{\mu\nu}&= \T^{\rm EM}_{\mu\nu} +\langle T^{\rm F}_{\mu\nu}\rangle\\
\nabla_{\mu} F^{\mu\nu} &= \langle J^{\nu}\rangle\\
\Gamma^{\mu} D_{\mu} \psi  &=m\,\psi.
\end{split}
\label{bulkeom}
\end{align}
Here the gauge field energy-momentum tensor $T_{\mu\nu}^{\rm {EM}}$ and the spinor energy-momentum tensor $T_{\mu\nu}^{\rm {F}}$ are given as operators by
\begin{align}
\begin{split}
T^{\rm EM}_{\mu\nu} &=  F_{\mu \rho} F_{\nu}^{\,\rho} - \frac{1}{4} g_{\mu\nu} F^2\\
T^{\rm F}_{\mu\nu} & = \frac{i \beta^2}{2} \overline{\psi} \Gamma_{(\mu}D_{\nu)}\psi + h.c.\label{FermiStressTensor},
\end{split}
\end{align}
while the spinor current operator is
\be
\label{current}
J^{\mu} = \beta^2 \overline{\psi} \Gamma^{\mu} \psi.
\ee
The expectation values in these expressions are taken in a ``bulk Fermi sea'', in which all the bulk fermionic modes with energy less than the chemical potential are populated. Computing these expectation values requires us to specify the single-particle wavefunctions of each of the occupied fermion states. We do this in a WKB approximation. We will construct the sea explictly and compute the expectation values of the current and the energy-momentum tensor operator in detail in section \ref{section-sea}. 

To solve the Einstein-Maxwell-Dirac system \eqref{bulkeom}, we take the following ansatz for the metric and the gauge field: 
\begin{align}
\begin{split}
ds^2 &= -f(z) dt^2 + \frac{1}{z^2}(dx^2 + dy^2) + g(z) dz^2,\\
A &= -h(z) dt - B x dy.
\end{split}
\label{bulkansatz}
\end{align}
We have taken Landau gauge for the electromagnetic potential.\footnote{In principle one could have allowed for $B = B(z)$. However, this gives an $x-z$ component to the stress tensor proportional to $B'(z)$,  which is inconsistent with our metric ansatz, so we must have $B = constant$.} In our coordinates, the boundary of the spacetime is at $z \to 0$, so the $B$ field is oriented from the IR out to the boundary. As for the asymptotic boundary conditions, we take the metric to be asymptotically AdS$_4$, i.e. $f,g \rightarrow 1/z^2$ as $z \rightarrow 0$. The gauge field satisfies $\underline{A}_t \rightarrow -\mu_{bdy}$ and $\underline{A}_{y} \rightarrow B_{bdy} x$, where these boundary quantities can be again rescaled and made dimensionless
\be
\mu_{bdy} = \frac{\gamma}{L} \mu, \ \ B_{bdy} = \frac{\gamma}{L^2} B
\label{bdyparams}
\ee
so that $h(z) \to \mu$ as $z \to 0$. Meanwhile, in the deep infrared we assume the existence of a horizon which carries some electric and magnetic charge; we describe this in detail in section \ref{section-background}, where we explicitly find the gravitational backgrounds.


Given the approximations described above, we will see that the stress tensor of the fermions takes the form of an anisotropic perfect fluid,
\be
\braket{T_{\mu\nu}^{\rm {F}}} = (\hat{\rho} +\hat{p}_\perp)u_\mu u_\nu + \hat{p}_\perp g_{\mu\nu}+(\hat{p}_z-\hat{p}_\perp)\chi_\mu\chi_\nu,
\label{stresstensor}
\ee
where the local thermodynamic quantities $\hat{\rho},\hat{p}_{\perp}, \hat{p}_z$ depend only on the holographic coordinate $z$ and on the external thermodynamic parameters $\mu, B,$ and $T$. Here $u^{\mu}$ is the unit vector along $\partial_t$ and $\chi^{\mu}$ the unit vector along $\partial_z$. We give the detailed form of the thermodynamic functions below. This form of the stress tensor means that the Einstein-Maxwell equations reduce to a coupled set of ODEs along the $z$ direction, which can be solved by numerical integration between the boundary and the horizon. This is the essential task of this paper.

\section{The bulk Fermi sea}
\label{section-sea}

\subsection{Definition of the sea}

As operators the bulk fermion energy-momentum tensor and current are given by the expressions in (\ref{FermiStressTensor}) and (\ref{current}). To compute the expectation values of these operators one should define a bulk state. We are after star-like gravitational solutions where the Dirac field has a large number of bound single-particle states, whose stress energy make up the star. In such a state it is natural to compute the current expectation values by summing up the contributions of all single-particle states with local energy $\hat{\omega}(z) = \omega/\sqrt{f(z)}$ less than the local chemical potential $\hat{h}(z) = h(z)/\sqrt{f(z)}$. Here $\omega$ is a rescaled energy $\omega = \gamma^{-1} \Omega$.

To proceed, we begin by writing the Dirac field in second quantization
\be
\label{modeexpansion}
\begin{split}
\psi(x) & = \sum_{\alpha} e^{-i \Omega_{\alpha} t} u_{\alpha}(\mathbf{x}) a_{\alpha} + e^{i \Omega_{\alpha} t}v_{\alpha}(\mathbf{x}) b_{\alpha}^{\dagger},\\
\psi^{\dagger}(x) & =  \sum_{\alpha} e^{-i \Omega_{\alpha} t} v^{\dagger}_{\alpha}(\mathbf{x}) b_{\alpha} + e^{i \Omega_{\alpha} t} u^{\dagger}_{\alpha}(\mathbf{x}) a^{\dagger}_{\alpha}.
\end{split}
\ee
The single-particle wavefunctions $u_{\alpha}(\mb{x})$ will be given explicitly later. Using (\ref{modeexpansion}), one easily obtains the tangent frame expressions
\begin{align}
\langle J^{\hat{a}}\rangle &= \beta^2 \langle  \overline{\psi} \Gamma^{\hat{a}} \psi \rangle= \frac{\beta^2}{\sqrt{f}} \sum_{\rm sea} u_{\alpha}^{\dagger} \Gamma^{\hat{t}} \Gamma^{\hat{a}} u_{\alpha},
\label{currentcomponents}
\end{align}
where the sum over the sea means a sum over local frequencies $0 \leq \hat{\omega}_{\alpha} \leq \hat{h}$. Similarly, for the expectation value of the stress tensor, one obtains the following expressions
\begin{align}
\begin{split}
\braket{T^{t}_{\ t}} &=\frac{\beta^2}{2\sqrt{f}} \sum_{\rm sea} u_{\alpha}^{\dagger} \Big( \hat{\omega}_{\alpha} - \hat{h} + \frac{i}{4\gamma} \omega_{\hat{t},\hat{a}\hat{b}} \sigma^{\hat{a} \hat{b}} \Big) u_{\alpha} +{\rm h.c.}  \\
\braket{T^{x}_{\ x}} & = -\frac{\beta^2}{2\sqrt{f}}  \sum_{\rm sea} u_{\alpha}^{\dagger} \Gamma^{\hat{t}} \Gamma^{\hat{x}} \Big(p_{\hat{x}} - \frac{i}{4\gamma} \omega_{\hat{x},\hat{a}\hat{b}} \sigma^{\hat{a} \hat{b}}\Big) u_{\alpha} + {\rm h.c.} \\
\braket{T^{y}_{\ y}} & = -\frac{\beta^2}{2\sqrt{f}}  \sum_{\rm sea} u_{\alpha}^{\dagger} \Gamma^{\hat{t}} \Gamma^{\hat{y}} \Big(p_{\hat{y}} - \hat{B}\hat{x} - \frac{i}{4\gamma} \omega_{\hat{y},\hat{a}\hat{b}} \sigma^{\hat{a} \hat{b}}\Big) u_{\alpha} + {\rm h.c.} \\
\braket{T^{z}_{\ z}} & = -\frac{\beta^2}{2\sqrt{f}}  \sum_{\rm sea} u_{\alpha}^{\dagger} \Gamma^{\hat{t}} \Gamma^{\hat{z}} p_{\hat{z}}  u_{\alpha} + {\rm h.c.}
\end{split}
\label{stresscomponents}
\end{align}
Here the momentum is $P_{\mu} = -i \partial_{\mu}$ as usual, and we defined the rescaled quantities
\begin{align}
\begin{split}
\hat{\omega} & = \frac{\hat{\Omega}}{\gamma}, \ \ p_{\hat{i}} = \frac{P_{\hat{i}}}{\gamma}, \ \ \hat{x} = e^{\hat{x}}_x x = \frac{x}{z}, \\
\hat{B} & = e^{x}_{\hat{x}} e^{y}_{\hat{y}} B = z^2 B = z^2 \gamma b. 
\end{split}
\end{align}

To proceed further, we need to find the single-particle wavefunctions $u_{\alpha}(\mathbf{x})$. This means we need to work out a complete set of solutions of the Dirac equation. Given the general ansatz \eqref{bulkansatz} for the metric and the gauge field, finding an analytic solution for the wavefunctions does not seem to be possible. However, analytic control over the solution can be obtained by performing a WKB analysis of the Dirac equation along the radial direction, as a systematic expansion in the small parameter $1/\gamma \ll 1$. We now turn to this task.

\subsection{WKB wavefunctions}
\label{wkbsection}

Consider the Dirac equation \eqref{bulkeom}. We can eliminate the spin connection terms by rescaling the single-particle wavefunctions $\psi = F \phi$ with $F = [-({\rm det}\,g) g^{zz}]^{-1/4} = z f^{-1/4}$. Using the time-independence of the metric, we can then rewrite the Dirac equation as a time-independent Schr\"{o}dinger problem
\be
\label{diraceqn3}
\hat{H} \phi = \hat{E} \phi, \ \ \hat{E} = \hat{\omega} - \hat{h},
\ee
where the local Hamiltonian is given by
\be
\hat{H} = -\Gamma^{\hat{t}} \left[ \Gamma^{\hat{i}} \left(p_{\hat{i}} - A_{\hat{i}} \right) + i m \right].
\ee
We would like to find a complete set of solutions, including the dispersion relation for $\omega$. One can write things in a transparent fashion. Define the ``magnetic algebra''
\begin{align}
\begin{split}
\hat{x} & = \frac{1}{\sqrt{2 \gamma \hat{B}}} \left( A + A^{\dagger} - \sqrt{\frac{2 \gamma}{\hat{B}}} p_{\hat{y}} \right) \\
p_{\hat{x}} & = -i \sqrt{\frac{\hat{B}}{2 \gamma}} \left(A - A^{\dagger} \right).
\end{split}
\label{magneticalgebra}
\end{align}
We have that $[ A, A^{\dagger} ] = 1$ on functions of $x$. In terms of these operators, the Hamiltonian is
\be
\hat{H} = \begin{pmatrix} m & \Oi \\ - \Oi & -m \end{pmatrix}
\ee
where
\be
\Oi = -\sqrt{\frac{\hat{b}}{2}} \left[ (A - A^{\dagger}) \sigma^{\hat{x}} + i (A + A^{\dagger}) \sigma^{\hat{y}} \right] - i p_{\hat{z}} \sigma^{\hat{z}}.
\ee

Solving the Dirac equation directly is impossible given our general metric ansatz. The implicit $z$-dependence of all these terms coming from the veilbeins is what makes things difficult. However, things are tractable in a WKB approach along this axis. Let us take the ansatz (justified shortly)
\be
\phi(\mb{x}) = X(x,z) Y(y,z) Z(z),
\ee
where the $z$-dependence in $X,Y$ is taken to be much slower than that of $Z$. In particular, we assume that the radial wavefunctions behave as
\be
Z \sim e^{i \gamma S(z)}
\ee
where $S(z)$ is a complex-valued function that we will solve for in the following, and recall that we have $\gamma \gg 1$. The WKB approximation is that we take the $z$-dependence of everything in the Dirac equation (\ie the metric and gauge field components) to be slow compared to the phase appearing here. This means that we can, to lowest order, commute $z$-derivatives past the veilbeins and terms involving the gauge field. To be precise, we assume that
\be
|\partial_{z} h|, \  |\partial_{z} e^{\mu}_{\hat{a}}| \ll \gamma |\partial_{z} S|, \ \ | \partial_{z}^2 S(z) | \ll \gamma | \partial_z S(z) |^2.
\ee

In this approximation, we can work with the much simpler problem of the square of the Dirac equation. This is because given a solution $\chi$ to the equation
\be
\hat{H}^2 \chi = \hat{E}^2 \chi
\ee
then the function $\phi = (\hat{H} + \hat{E}) \chi$ will solve \eqref{diraceqn3}, to lowest order in the WKB approximation. We have that
\be
\hat{H}^2 = m^2 + 2 \hat{b} \left( n + \frac{1}{2} + \frac{1}{2} \hat{\Sigma} \right) + p_{\hat{z}}^2.
\ee
where $n = A^{\dagger} A$ and $\hat{\Sigma} = -i \Gamma^{\hat{x}} \Gamma^{\hat{y}} = \sigma^{\hat{z}} \otimes \mb{1}_{2 \times 2}$ is the local spin operator. This is clearly block-diagonal and each $2 \times 2$ block is identical. The solution in a particular block can then be taken as
\be
\label{diracsquaredsolns}
\chi = N \psi_n(\hat{x} + p_{\hat{y}}/\hat{B}) e^{i P_y y} Z(z) \zeta_{\pm},
\ee
where $\zeta_{\pm}$ is an eigenstate of $\sigma^{\hat{z}}$, and $\psi_n$ is the $n$th state of a harmonic oscillator with frequency $\hat{b}$.

All that is left to do is to work out the $z$-dependence. Given \eqref{diracsquaredsolns}, one finds the WKB problem for the holographic direction is simply 
\begin{align}
\label{radialeqn}
\partial_{\hat{z}}^2 Z(z) = \gamma^{2} V(z) Z(z)
\end{align}
where the effective Schr\"{o}dinger potential is given by
\be
\label{potential}
V = V_{n,\pm}(z) = m^2 + 2 \hat{b} \Big( n + \frac{1}{2} \pm \frac{1}{2} \Big) - (\hat{\omega}- \hat{h})^2.
\ee
This is just a one-dimensional Schr\"{o}dinger problem at zero energy. Note that $V_{n-} = V_{(n-1)+}$. To lowest order, the WKB solution is
\begin{align}
\begin{split}
Z(z) & = e^{\pm i \gamma S(z)},\\
S(z) & = \int^{z} dz' \ \hat{q}(z'), \ \ q(z) = \sqrt{-V(z)}.
\end{split}
\label{radialsoln}
\end{align}
Here, the sign choice in the exponent is for incoming or outgoing radial modes, and $\hat{q} = e^{z}_{\hat{z}} q$.

\begin{figure}
\includegraphics[scale=0.63]{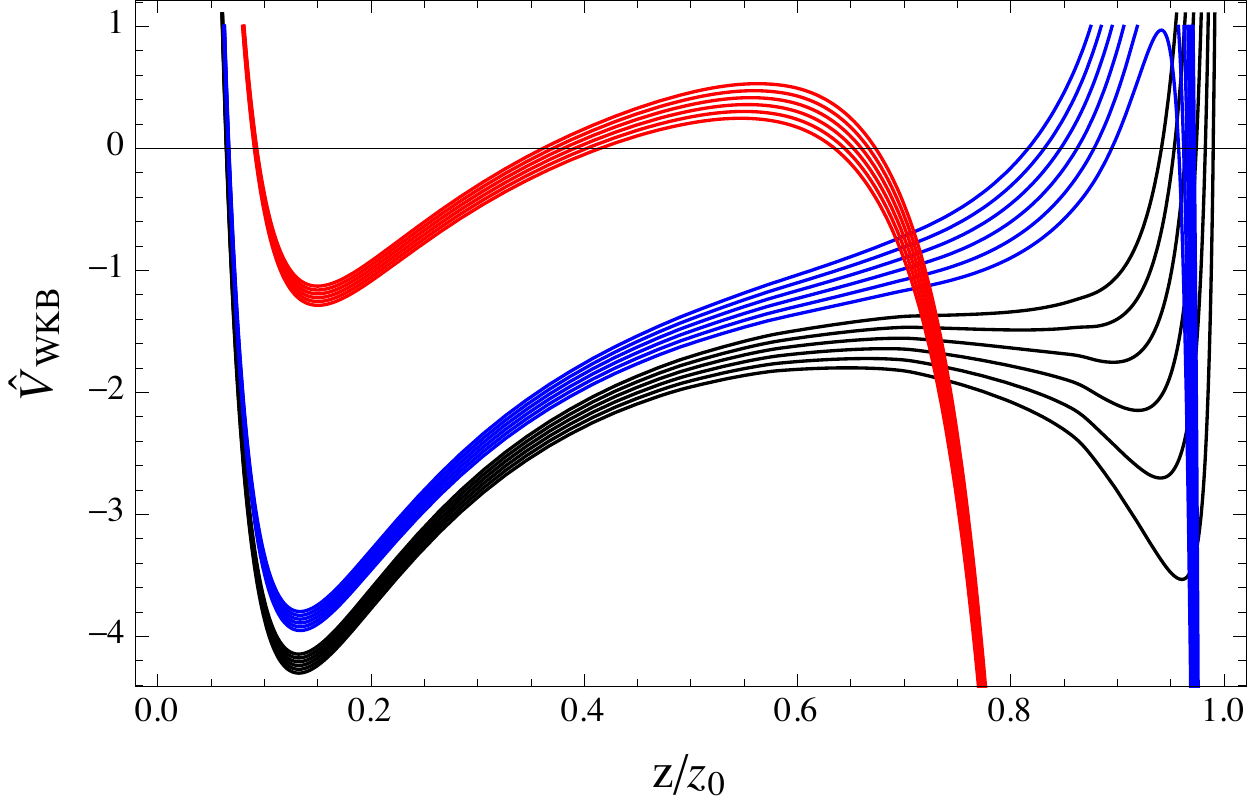}
\caption{Local effective WKB potential $\hat{V} = g V$, with $V$ given in \eqref{potential}, for some low-lying Landau levels $n = 0,\ldots,5$ in the bulk. The different colors denote different frequencies: black is $\omega=0$, blue is $0 < \omega \ll 1$, and red is $\omega \sim 1$. This is evaluated in a backreacted star background of section \ref{section-background}, with $T/\mu \ll B/\mu^2 \ll 1$. Here $z_0$ is the horizon.}
\label{wkbplotsomega=0}
\end{figure}

The kind of gravitational solutions we will be interested in are such that the effective potential (\ref{potential}) will admit bound states in a finite region of $z$; see for example figure \ref{wkbplotsomega=0}. The bound states supported by this potential are the states that make up the star. The WKB momenta allowed in this compact region will form a discrete (if very dense) set, determined by a Bohr-Sommerfeld condition of the form $\int_{z_0}^{z_1} q dz = N$. Here $N$ is an integer and $0 < z_0 < z_1$ label the classical turning points of (\ref{potential}). These turning points are just the zeros of the potential since (\ref{radialeqn}) is a zero-energy Schr\"{o}dinger equation. 

We can read off the (local) energy spectrum. We label the eigenstates of $\hat{H}^2$, i.e. the functions \eqref{diracsquaredsolns}, as $\chi = \chi_{pqn\pm}$ in the notation of the previous few paragraphs. Here $p$ indexes the momenta along the $y$-axis. These states have local energy
\be
\hat{E}_{qn\pm}^2 = m^2 + \hat{q}^2 + 2 \hat{b} \Big( n + \frac{1}{2} \pm \frac{1}{2} \Big).
\ee
Note that this is completely degenerate in the momentum $p$ along the $y$-axis. Moreover, the state at Landau level $n$ with spin down is degenerate with that at level $n-1$ and spin up, and we write
\be
\label{spectrumwoo}
\hat{E}_{qn}^2 = m^2 + \hat{q}^2 + 2 \hat{b} n = \hat{E}^2_{qn-} = \hat{E}^2_{q(n-1)+}.
\ee
Including antiparticle states, each energy has fourfold degeneracy except the ground state $n=0, \sigma^{\hat{z}} = -1$ which is only doubly-degenerate. In flat spacetime, these degeneracies can be accounted for by an internal $\mathcal{N} = 2$ SUSY algebra \cite{wittensusy,Haymaker:1985us}, and it would be interesting to understand how this works in AdS.

Finally, we can act with $\hat{H} + \hat{E}$ on the wavefunctions \eqref{diracsquaredsolns} to get a complete set of solutions to the Dirac equation proper, \eqref{diraceqn3}. For the particle states we can begin with the multiplet 
\be
\chi^1_{pqn} = \begin{pmatrix} \chi_{pqn-} \\ 0 \end{pmatrix}, \ \ \chi^2_{pqn} = \begin{pmatrix} \chi_{pq(n-1)+} \\ 0 \end{pmatrix}
\ee
both members with $\hat{E} = \hat{E}_{qn}$. Then, acting with $\hat{H} + \hat{E}$, we obtain the single-particle states
\be
\phi^{i}_{pqn} = X_{n}^{i} Y_{p} Z_{qn}
\ee
with $Y_{p} = e^{i \gamma p y}/\sqrt{\ell}$ a box-normalized plane wave, the spin-oscillator wavefunctions
\begin{align}
\begin{split}
X_{qn}^{1} & = N_X \begin{pmatrix} (\hat{E}_{qn} + m) \chi_{pqn-} \\ \sqrt{2\hat{b}n} \chi_{pq(n-1)+} - i \hat{q} \chi_{pqn-} \end{pmatrix} \\
X^{2}_{qn} & = N_X \begin{pmatrix} (\hat{E}_{qn} + m) \chi_{pq(n-1)+} \\ -\sqrt{2\hat{b}n} \chi_{pqn-} + i \hat{q} \chi_{pq(n-1)+} \end{pmatrix},
\end{split}
\label{positivestates}
\end{align}
and the radial wavefunctions $Z_{qn} = Z_{qn-} = Z_{q(n-1)+}$ given by \eqref{radialsoln}. One can obtain quite similar formulae for a multiplet of antiparticle states, starting with the $\chi$'s in the lower block rather than the upper block. Working in a transverse box of coordinate area $\ell^2$, one finds that normalizing the spin-oscillator wavefunctions 
\be
\delta_{nm} \delta_{ij} = \int dx X_{n}^{i \dagger} X_{m}^j
\ee
sets $|N_X|^2  = \left[ 2 \hat{E}_{qn}(\hat{E}_{qn} + m) \right]^{-1}$.

\subsection{WKB fermionic currents}

We need to work out the fermionic contribution to the stress energy. Continuing to regulate the computations by considering a finite coordinate area $\ell^2$ on the $x-y$ plane, the degeneracy of single-particle states is finite. 

When we sum over the local Fermi sea as in \eqref{currentcomponents}, we sum over the contributions from the states with local energy below the local chemical potential, $0 \leq \hat{E}_{qn} \leq \hat{h}$.  In terms of the quantum numbers $n,q$, this means states with
\be
\left| \hat{q} \right| \leq \hat{q}_n := \hat{h}^2 - \hat{m}_n^2, \ \  n \leq \hat{n}_F := \frac{\Delta \hat{m}^2}{2 \hat{b}}
\ee
where we defined
\be
\hat{m}_n^2 = m^2 + 2 \hat{b} n, \ \ \Delta \hat{m}^2 = \hat{h}^2-m^2.
\ee
Note that these quantities are local, and in particular are $z$-dependent. In regions where the fermion mass $m > \hat{h}$, no states can be occupied, and the fermion stress tensor elements are all zero. With these conventions, the sum over the sea means
\be
\sum_{\text{sea}} = \frac{\hat{b} \ell^2}{2\pi} \sum_{i=1,2} \sum_{n=0}^{\hat{n}_F} \sum_{\hat{q} = -\hat{q}_n}^{\hat{q}_n}.
\ee 
Here the prefactor is from the degeneracy on the $x-y$ plane: we need $| \hat{p} | \leq \hat{b} \ell/2$ so that the shifted oscillator is still within the transverse box, and there are $\hat{b} \ell^2/2\pi$ such states. The sum over $n$ is doubly degenerate except for $n=0$, so at each Landau level $n$ the degeneracy is $\hat{g}_n = \ell^2 \hat{d}_n$ with
\be
\hat{d}_n = \frac{\hat{b}}{2\pi} (2-\delta_{n,0}).
\ee

Let us begin with the current. Starting with \eqref{currentcomponents} and computing directly with the wavefunctions \eqref{positivestates}, making use of the translational invariance in the $x-y$ plane, we have for the local charge density
\be
\begin{split}
\hat{\sigma} = \langle J^{\hat{t}}\rangle & =  \beta^2 \frac{F^2}{\sqrt{f}} \sum_{n=0}^{\hat{n}_F} \sum_{\hat{q} = -\hat{q}_n}^{\hat{q}_n} \hat{d}_n \left| Z_{qn}(z) \right|^2.
\end{split}
\label{densityads}
\ee
As for the spacelike components, it is easy to check by direct computation that they all vanish:\footnote{The fact that $J^z = 0$ is a reflection of the WKB approximation, in which we neglected the electric field $E^z \sim h'(z)$.}
\be
\langle J^{\hat{x}}\rangle = \langle J^{\hat{y}}\rangle = \langle J^{\hat{z}} \rangle= 0.
\ee

On to the stress tensor \eqref{stresscomponents}. In the WKB approximation, the spin term in the stress tensor is suppressed by a factor of $\gamma^{-1}$, so we drop it.\footnote{Physically this means we are ignoring the backreaction of gravitational spin-orbit coupling.} Using the same techniques as for the current, and using the magnetic algebra \eqref{magneticalgebra}, one obtains
\begin{align}
\begin{split}
\langle T_{\hat{t}\hat{t}} \rangle & = \beta^2 \frac{F^2}{\sqrt{f}} \sum_{n=0}^{\hat{n}_F} \sum_{\hat{q} = -\hat{q}_n}^{\hat{q}_n} \hat{E}_{qn} \hat{d}_n \left| Z_{qn}(z) \right|^2 \\
\langle T_{\hat{x}\hat{x}} \rangle & = \beta^2 \frac{F^2}{\sqrt{f}} \sum_{n=0}^{\hat{n}_F} \sum_{\hat{q} = -\hat{q}_n}^{\hat{q}_n} \hat{d}_n \frac{\hat{b} n}{\hat{E}_{qn}} \left| Z_{qn}(z) \right|^2 = \langle T_{\hat{y}\hat{y}} \rangle \\
\langle T_{\hat{z}\hat{z}} \rangle & = \beta^2 \frac{F^2}{\sqrt{f}} \sum_{n=0}^{\hat{n}_F} \sum_{\hat{q} = -\hat{q}_n}^{\hat{q}_n} \hat{d}_n \frac{\hat{q}^2}{\hat{E}_{qn}} \left| Z_{qn}(z) \right|^2 \\
\end{split}
\end{align}

We are interested in the situation in which a large number of states are occupied, so that in particular we can approximate the sum over the WKB momentum as an integral. Moreover, we expect that in this limit, the fermions can be treated in a fluid approximation, and their equation of state should be the same as in flat space, except with the external parameters evaluated in the local tangent frame. Comparing \eqref{densityads} to \eqref{currenttimecomponent} then suggests that we identify the local density of states, following \cite{Ruffini:1969qy}, as\footnote{There is some ambiguity in whether the overall factor is absorbed into a radial wavefunction normalization or into this definition of the density of states. For our purposes it does not matter how we do it; even when we compute the boundary CFT correlators the overall wavefunction normalizations will drop out \cite{Son:2002sd}.}
\be
\sum_{\hat{q} = -\hat{q}_n}^{\hat{q}_n} \frac{F^2}{\sqrt{f}} \left| Z_{qn}(z) \right|^2 \to \int_{-\hat{q}_n}^{\hat{q}_n} d\hat{q}.
\ee
Making this identification in the stress tensor and performing the $d\hat{q}$ integrals, one obtains
\begin{align}
\begin{split}
\langle J^{\hat{t}}\rangle & = 2 \beta^2 \sum_{n=0}^{\hat{n}_F} \hat{d}_n \hat{q}_n \\
\langle T_{\hat{t} \hat{t}} \rangle & = 2 \beta^2 \sum_n^{\hat{n}_F} \hat{d}_n \left[ \hat{h} \hat{q}_n + \hat{m}_n^2 \ln \left( \frac{\hat{q}_n + \hat{h}}{\hat{m}_n} \right) \right] \\
\langle T_{\hat{x} \hat{x}} \rangle & = 4 \beta^2 \sum_n^{\hat{n}_F} \hat{d}_n \left[ 2 \hat{b} n\ln \left( \frac{\hat{q}_n + \hat{h}}{\hat{m}_n} \right) \right] \\
\langle T_{\hat{z} \hat{z}} \rangle & = 4 \beta^2 \sum_n^{\hat{n}_F} \hat{d}_n \left[ \hat{h} \hat{q}_n - \hat{m}_n^2 \ln \left( \frac{\hat{q}_n + \hat{h}}{\hat{m}_n} \right) \right].
\end{split}
\end{align}

At this stage, it is clear that we have reduced the fermion stress tensor to that of an axisymmetric perfect fluid \eqref{stresstensor}. The local thermodynamic quantities appearing in \eqref{stresstensor} are given by the local stress tensor components given above, that is $T_{\mu\nu} = e^{\hat{a}}_{\mu} e^{\hat{b}}_{\nu} T_{\hat{a} \hat{b}}$. 

These expressions still involve a sum over Landau levels, which in our approximations will involve a summation over $\hat{n}_F \sim \Oi(\gamma)$ terms. These expressions are exact to lowest order in $\gamma$, but as a final approximation, we can expand the thermodynamics as a Poisson series in $\hat{n}_F \sim \hat{b}^{-1}$ (see appendix \ref{dhvaappendix} for the derivation in flat space). This expresses the equation of state as that of a free, degenerate Fermi gas at $T=0$ plus contributions oscillatory in $\hat{b}^{-1}$ and with amplitude supressed by a power of $\hat{b}/\hat{h}^2 \sim \Oi(\gamma^{-1})$. The oscillatory parts encode the famous de Haas-van Alphen oscillations. Expanding, one obtains the zeroth-order thermodynamics
\begin{align}
\begin{split}
\hat{\sigma}_0 &= \frac{\beta^2 \Delta \hat{m}^3}{3\pi^2} \\
\hat{\rho}_0 & = \frac{\beta^2}{8\pi^2}\left[\hat{h}(2\hat{h}^2-m^2) \Delta \hat{m} - m^4\ln{ \left(\frac{\hat{h} + \Delta \hat{m}}{m}\right)} \right] \\
p_0&=\frac{\beta^2}{24\pi^2} \left[\hat{h}(2\hat{h}^2-5m^2) \Delta \hat{m} +3 \ln{\left(\frac{\hat{h} + \Delta \hat{m}}{m} \right)}  \right].
\end{split}
\end{align}
Since the equation of state at this order is independent of $\hat{b}$, the local pressure $\hat{p}_0$ is isotropic. The leading order correction in the magnetic field is
\begin{align}
\begin{split}
\hat{\sigma} &=\hat{\sigma}_0+\frac{\beta^2 \hat{b}^{3/2}}{\pi^3} \sum_{k=1}^{\infty}\frac{1}{k^{3/2} }\sin\Big(2\pi k\hat{n}_F -\frac{\pi}{4}\Big) \\
\hat{\rho} &= \hat{\rho}_0+\frac{\beta^2 \hat{h} \hat{b}^{3/2}}{\pi^3} \sum_{k=1}^{\infty}\frac{1}{k^{3/2} }\sin\Big(2\pi k\hat{n}_F -\frac{\pi}{4}\Big) \\
\hat{p}_z &=\hat{p}_{0}-\frac{\beta^2 \hat{b}^{5/2}}{4\pi^4\hat{h}} \sum_{k=1}^{\infty}\frac{1}{k^{5/2} }\cos\Big(2\pi k\hat{n}_F -\frac{\pi}{4}\Big) \\
\hat{p}_\perp &= \hat{p}_{0}+\frac{\beta^2 \hat{b}^{3/2}}{2\pi^3\hat{h}} \Delta \hat{m}^2 \sum_{k=1}^{\infty}\frac{1}{k^{5/2} }\sin\Big(2\pi k\hat{n}_F -\frac{\pi}{4}\Big).
\end{split}
\label{fullthermovars}
\end{align}

In these equations, we have a zero-temperature fluid. This is automatic from our definition of the bulk state in that we simply populated a zero-temperature bulk Fermi sea. However, the horizons we will assume in the infrared generically carry a temperature; we are working in the limit in which we ignore the effect of the Hawking radiation on the fluid. To justify this, note that the local temperature felt by the fermions scales like $T_{local}/\mu_{local} \sim \Oi(\gamma^{-1})$, where $\mu_{local} = \gamma \hat{h}$ is the unscaled, physical chemical potential, and we assume $T/\mu \ll 1$ (see also \cite{Hartnoll:2010ik}).

\section{Dyon star backgrounds}
\label{section-background}

\begin{figure}
\includegraphics[scale=0.6]{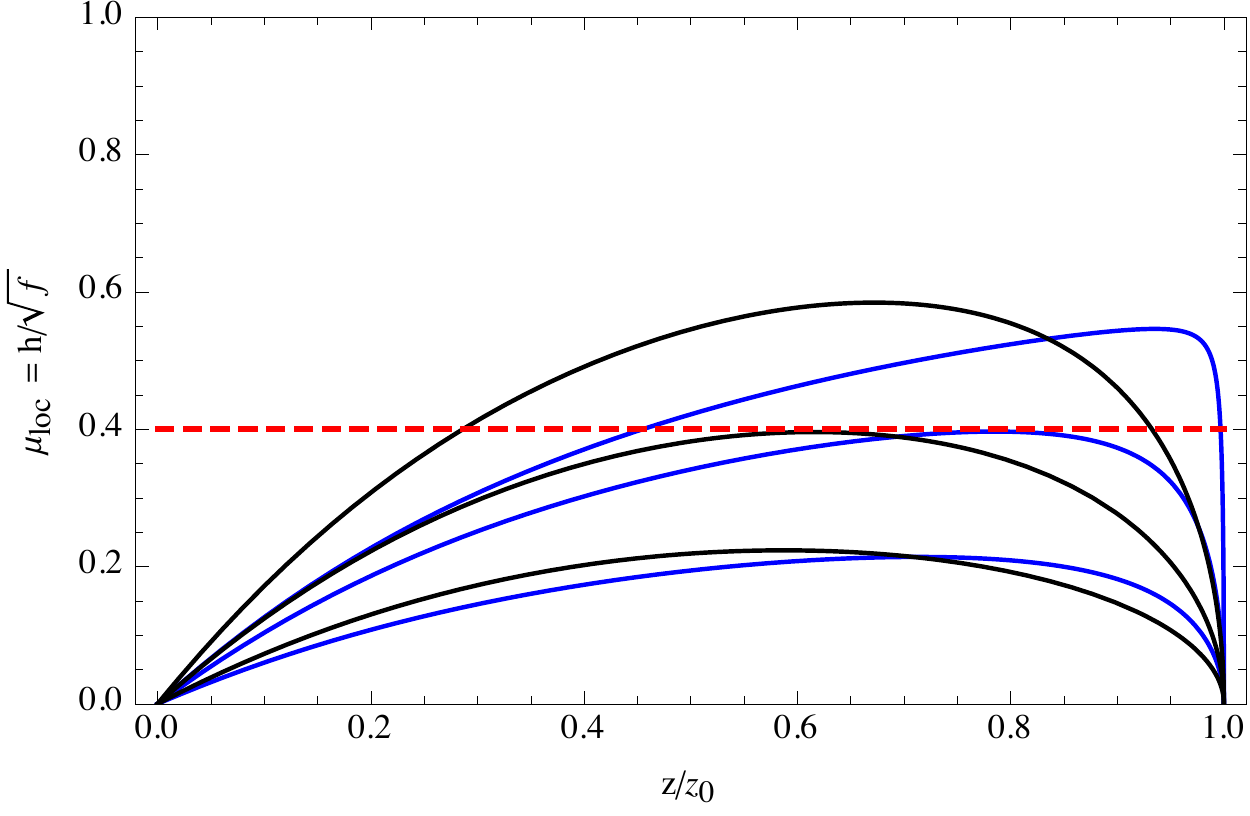}
\caption{Local chemical potential $\hat{h}(z) = h(z)/\sqrt{f(z)}$ in the pure black brane background \eqref{AdS-RN} with $T > 0$. The red dashed line is the fixed fermion mass, black curves are the local chemical potential with $B = 0$ while blue have $B > 0$, and in both cases the horizon temperature is increasing from top to bottom.}
\label{localpotentialfig}
\end{figure}

Given the expression \eqref{stresstensor} for the stress-energy of the fluid and the ansatz \eqref{bulkansatz}, one finds that the Einstein equations \eqref{bulkeom} reduce to
\begin{align}
\begin{split}
0&=\frac{1}{z}\left(\frac{f'}{f}+\frac{g'}{g}+\frac{4}{z}\right) +\left(\hat p_z+\hat \rho\right) g,\\
0&=\frac{f'}{zf}-\frac{h'^2}{2f}+\left(3+\hat p_z-B^2 z^4\right)g-\frac{1}{z^2},\\
0&=h''+\frac{z}{2}\left(\hat p_z+\hat \rho\right)gh'-g\sqrt{f} \hat\sigma.
\end{split}
\label{finaleom}
\end{align}

There is by now a well-known algorithm \cite{Hartnoll:2010gu} for constructing asymptotically AdS stars satisfying these equations, which we will follow here. As explained earlier, to get a boundary temperature and magnetic field we assume the existence of an electrically and magnetically charged planar black brane deep in the IR, say at $z=z_0 > 0$.\footnote{These are the planar limits of the dyonic black holes discussed originally by Romans \cite{Romans:1991nq}, see also \cite{Hartnoll:2007ai}.} In such a pure black brane background with $T>0$, the local chemical potential $\hat{h}(z)$ is a convex function equal to zero on the boundary and horizon, see fig. \ref{localpotentialfig}. Thus for fixed fermion mass, one can lower the temperature or raise the boundary chemical potential such that $m \leq \hat{h}$ for some finite region $z_- < z < z_+$. This region will then have a finite density of fermion states, followed by another Reissner-Nordstrom solution in the exterior region from $z=z_-$ out to the boundary $z=0$. The physical parameters of the boundary theory can then be read off from the exterior solution.

The interior geometry and gauge field, from the horizon out to the boundary of the star $z_-$, are given by
\begin{align}
\begin{split}
f(z) & = \frac{1}{z^2} \left[1 - M_0 z^3 + (H^2 + Q_0^2) z^4 \right],  \\
g(z) & = \frac{1}{f(z) z^4}, \ \ h(z) = \mu_0 - \sqrt{2} Q_0 z, \ \ B = \sqrt{2} H.
\end{split}
\label{AdS-RN}
\end{align}
Here $Q_0$ and $H$ are, respectively, the electric and magnetic charges of the black hole, the horizon $z_0$ is determined by the first positive root of $f(z_0) = 0$, and $\mu_0$ is a constant. Regularity of the stress tensor requires $h(z_0) = 0$, that is $\mu_0 = \sqrt{2} Q_0 z_0$, and the mass of the black hole is given by $M_0 = z_0^{-3} \left[ 1 + (H^2 + Q_0^2)z_0^4 \right]$. Since no fermion states are occupied in this region, we have $\hat\sigma=\hat\rho=\hat p_z=\hat p_\perp=0$ and one can check easily that \eqref{AdS-RN} solves \eqref{finaleom}. 

Fix the fermion mass $m$, magnetic field $B$ and boundary chemical potential $\mu$. Starting with sufficiently high $T/\mu$, one has that $\hat{h} > m$ everywhere and the full solution is simply the black brane given above. As one cools the system, there will be a particular value $T = T_c$ and radial coordinate $z=z_c$ at which
\be
\hat{h}(z_c) = m, \ \ \partial_z \hat{h}(z_c) = 0, \ \ (T = T_c)
\ee
where the star is ``born''. Roughly speaking, one has an infinitesimal shell of fermions at $z=z_c$. Continuing to lower the temperature $T < T_c$, the width of the star increases monotonically, occupying the region between the two roots $z = z_{\pm}$ of
\be
\hat{h}(z_{\pm}) = m. \ \ \ (T < T_c)
\ee
In the region $z_- < z < z_+$, a finite density of fermions contributes to the stress-energy and backreacts on the geometry. The solution in this region must be obtained by numerically integrating \eqref{finaleom}, subject to continuity of the functions $f,g,h$ and $\partial_z h$ at $z=z_{\pm}$. 

Finally, for the exterior region $0 < z < z_-$, we again have zero fermion stress-energy, and so the metric and gauge fields will again be given by the Reisner-Nordstrom solution, although with some different parameters and an overall constant rescaling:
\begin{align}
\begin{split}
f(z) & = \frac{c^2}{z^2} \left[1 - M z^3 + (H^2 + Q^2) z^4 \right],  \\
g(z) & = \frac{c^2}{f(z) z^4}, \ \ h(z) = c( \mu - \sqrt{2} Q z) , \ \ B = \sqrt{2} H
\end{split}
\label{AdS-RN-exterior}
\end{align}
Note that we do not need to adjust the magnetic field, consistent with the fact that $B = constant$ by the symmetry of our metric ansatz. These four equations determine the exterior parameters $c,M,Q,\mu$. 

The temperature as measured from the boundary is given by
\be
T = \frac{1}{4\pi c} \left| \frac{df}{dz} \right|_{z=z_0}.
\ee
The factor $c$ is unity in the pure Reissner-Nordstrom solution; backreaction of the fermions sets $c \neq 1$. Given this equation, we can set $z_0 = 1$ without loss of generality. The interior solutions are then parametrized simply in terms of $\mu_0$ and $H$; by varying these parameters one can obtain any desired values of the ratios $B/\mu^2$ and $T/\mu$ in the exterior.

\begin{figure}
\includegraphics[scale=0.6]{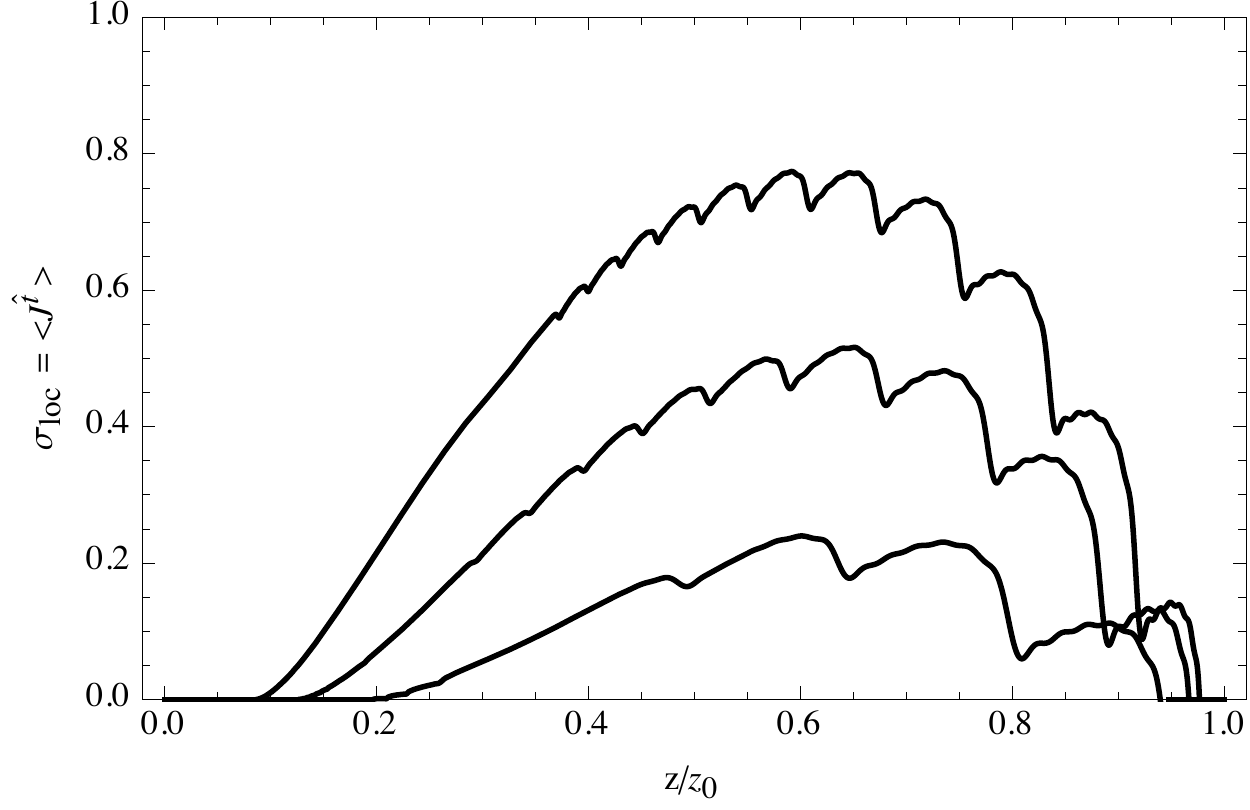}
\caption{Local charge density $\hat{\sigma}(z)$ of some prototypical dyon star solutions at $T > 0$, with $T/\mu$ increasing from top to bottom. One clearly sees the de Haas-van Alphen oscillations reflected in the stellar structure.}
\label{wavystar}
\end{figure}

In figure \ref{wavystar}, we plot the charge density of some solutions of this system as a function of $z$. The gross features of the star, determined by the $B$-independent part of the thermodynamics, are identical to those of the purely electron stars of \cite{Hartnoll:2010gu,Hartnoll:2010ik}. The first order correction in $B/\mu^2$ provides an elegant manifestation of the de Haas-van Alphen effect. As the local magnetic field $\hat{b} = \hat{b}(z)$ varies continuously with $z$, the thermodynamic variables \eqref{fullthermovars} are oscillatory in $z$, leading to periodic features in the stellar structure.

Finally, let us study the nature of the phase transition between the pure black brane and the star solutions as the external parameters are varied. As explained in detail in \cite{Hartnoll:2010gu,Hartnoll:2010ik}, the free energy of the bulk solution can be computed purely in terms of the boundary data,\footnote{The absence of a magnetic term can be understood from the falloffs of the gauge field: $A_y = B x$ exactly, and there is no subleading behavior as a function of $z$.}
\be
\Omega = M - \mu Q - s T.
\ee
To be precise, this is the free energy per unit transverse area, rescaled by $L/\gamma$ like all the other energies considered in this paper. Here $s$ is the rescaled entropy density of the black brane,
\be
s = 2 \pi \frac{L}{\kappa} = 2 \pi \frac{\gamma^2}{\beta}.
\ee

\begin{figure}
\includegraphics[scale=0.65]{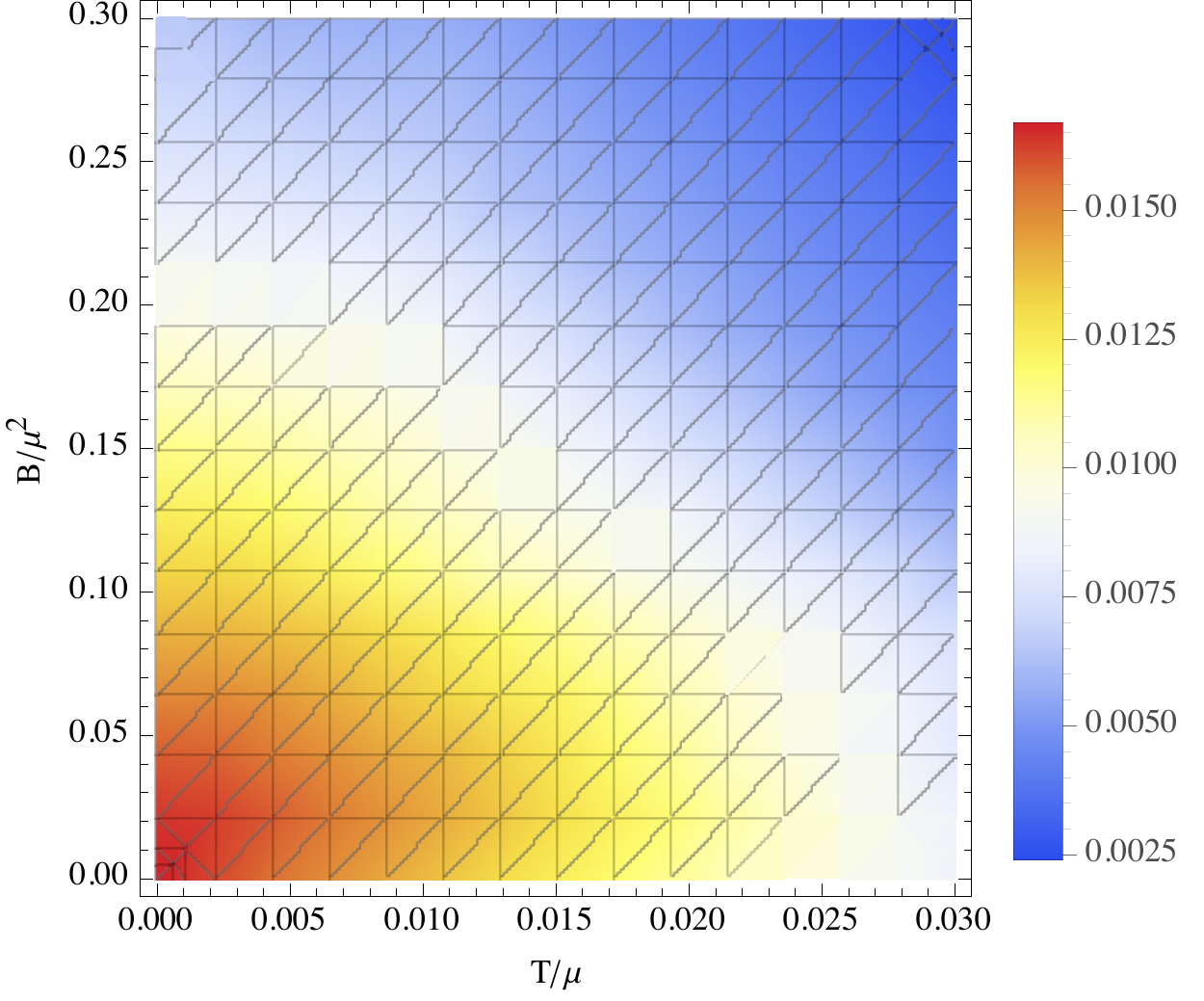} \\
\includegraphics[scale=0.65]{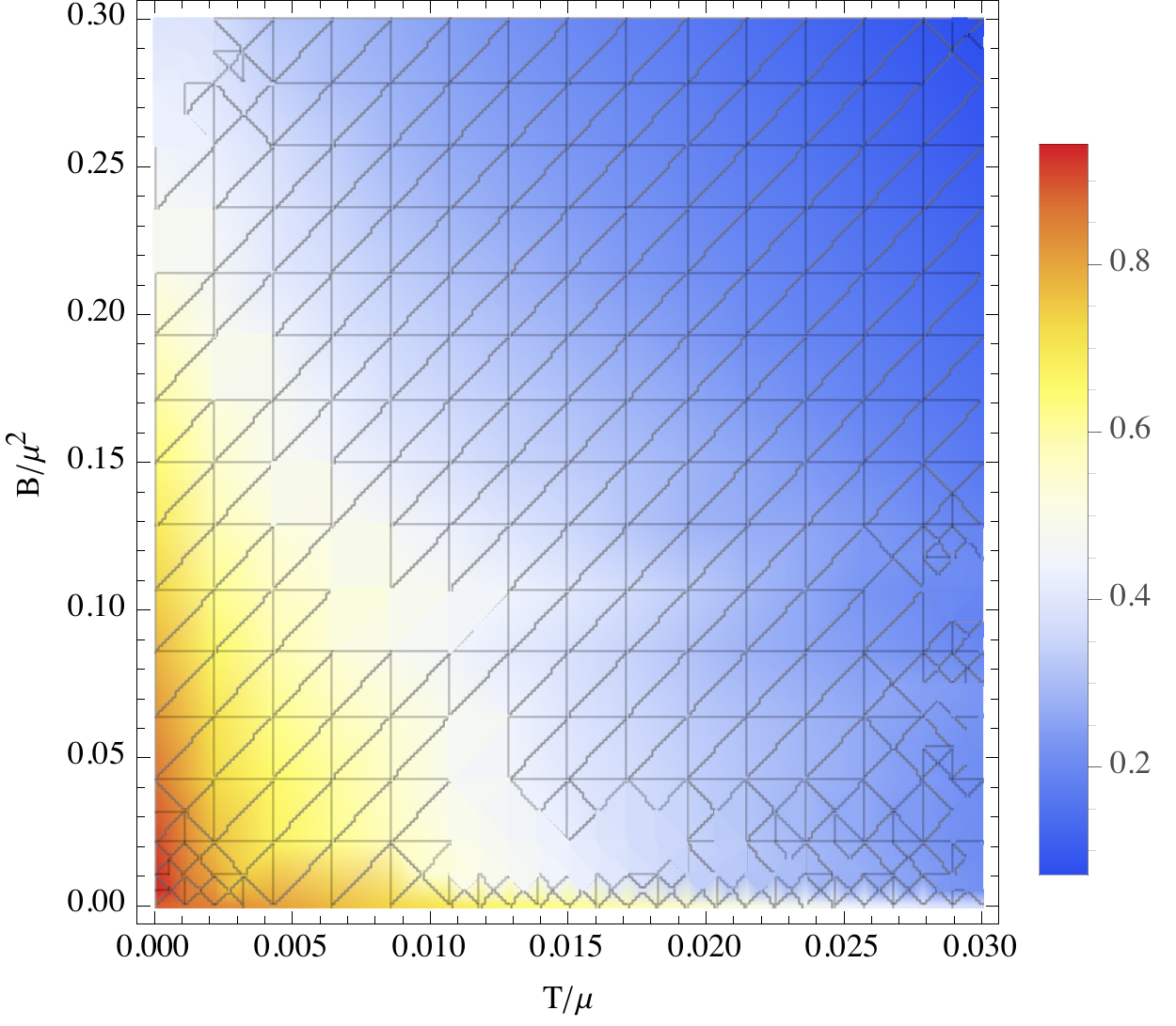}
\caption{Top: Comparison of the free energy $\Delta \Omega = \Omega_{BB} - \Omega_{star}$ between the star and pure black brane backgrounds. One sees easily that the stars are thermodynamically favored at sufficiently low $T/\mu$ and $B/\mu^2$. Bottom: Fraction of the bulk charge $Q_f$ due to bulk fermions.}
\label{freeenergypic}
\end{figure}

In figure \ref{freeenergypic}, we show the difference in free energies between the star and black brane solutions with equal external parameters. One can easily see that at low temperatures $T < T_c(B)$, the star solution is thermodynamically favored over the pure black brane. The phase transition is extremely soft; in \cite{Hartnoll:2010ik} it was argued to be third order, and the same arguments appear to apply here.

With the chemical potential and magnetic field kept fixed, one can consider the zero-temperature limit. In this limit, the black brane becomes extremal. The local chemical potential limits to a \emph{finite} value at the horizon, $\hat{h}(z_0) = \mu z_0/\sqrt{6}$, and then decreases monotonically outward to zero at the boundary. This would suggest that if we take the fermion mass sufficiently small with $\mu > 0$ fixed, one would have a finite density of populated fermions all the way down to the horizon. 

To check this idea, one can consider the fraction of the bulk electric charge carried by the fermions,
\be
Q_f = \frac{Q_{\psi}}{Q}, \ \ Q_{\psi} = \int dz\ \frac{\sqrt{g(z)}}{z^2} \hat{\sigma}(z).
\label{Qfdef}
\ee
As described in \cite{Hartnoll:2010ik}, in the purely electric case $B=0$, as one lowers $T/\mu \to 0$ one finds that $Q_f \to 1$. All the bulk charge is ejected from the brane into the bulk, leaving nothing to source the geometry behind the horizon, and one finds an emergent Lifshitz scaling geometry in the deep IR replacing the black brane. 

In our case, according to figure \ref{freeenergypic}, once a finite magnetic field is turned on, one has instead that $Q_f$ limits to a finite value less than one as $T/\mu \to 0$. This value decreases monotonically with increasing $B/\mu^2$. This means that the brane remains as part of the solution in the deep IR and continues to source both the bulk magnetic field and chemical potential; a finite fraction of the bulk charge remains behind the horizon.

\section{Conclusions} 

In this paper we have proposed a gravitational dual to systems of fermions strongly interacting via a $U(1)$ current, at finite density and temperature, subjected to an external, transverse magnetic field. Our gravitational solutions, thermodynamically favored over pure black hole backgrounds at low temperature, consist of a condensate of weakly-coupled bulk fermions hovering over an electrically and magnetically charged black brane in the IR. We have called these solutions dyon stars.

These dyon stars are solutions of Einstein gravity coupled only to a bulk $U(1)$ field weakly coupling a bulk fermion. This kind of minimal setup should be a robust approximation at low energies for a large class of field theories. 

We have given numerical star solutions for $T>0$ and $B > 0$, and argued for a scaling limit as $T \to 0$ in which part of the bulk charge is still carried by a brane in the deep IR. However, explicitly constructing the $T=0$ solutions is not possible in our setup, because the coordinate singularity at the horizon becomes too severe. Very near the horizon (for any $T$), various veilbeins will diverge and thus the WKB approximation becomes bad. One presumably needs to move to some coordinates in which the horizon is non-singular to give the $T=0$ construction.

One particularly interesting question would be to understand the relationship between the bulk ``Fermi sea'' state we have been using here, and the actual Fermi sea $\ket{\mu}$ of the dual CFT constructed by occupying the lowest-lying CFT states. In our treatment, following the usual fluid approximations, we simply neglected the stress-energy from the parts of the fermion wavefunctions where the fermion is unbound. It would be interesting to understand such contributions in detail.

The dynamics of strongly-coupled fermions at high density is of obvious and practical interest, and the addition of an external magnetic field is quite relevant for realistic lab systems. Holography continues to be a promising approach to such systems, and our work paves the way for the systematic study of CFTs with these features.

\acknowledgements{We thank Willy Fischler, Sean Hartnoll, and Joe Polchinski for conversations, and Rob Leigh for early collaboration. Our work at UT is supported by NSF grant PHY-1316033, and DC is supported by Templeton Foundation award 36838, by NSERC, and by the Pacific Institute for Theoretical Physics.}

\appendix

\section{Spin geometry}
\label{spingeo}
To define the Dirac action in curved spacetime requires one to specify the spin geometry of the spacetime. One first needs to define a veilbein, which we will take to be
\be
\label{veilbein}
e_{\hat{a}}^{\mu} = \begin{pmatrix}
	\frac{1}{L \sqrt{f(z)}} & & & \\
	 & \frac{z}{L} &  &  \\
	 &  & \frac{z}{L} &  \\
	 &  &  & \frac{1}{L \sqrt{g(z)}}  \end{pmatrix}.
\ee 
Here $\mu \in \{ t, x,y,z \}$ is the bulk spacetime index while $\hat{a} \in \{ \hat{t},\hat{x},\hat{y},\hat{z} \}$ is the bulk tangent frame index. As usual frame indices are raised and lowered with the flat metric $\eta_{\hat{a} \hat{b}} = g_{\mu\nu} e^{\mu}_{\hat{a}} e^{\nu}_{\hat{b}}$. We will often define ``local'' or ``tangent frame'' quantities by exchanging spacetime for frame indices; for example, the local energy and momenta are defined by
\be
p_{\mu} = (E,p_i) \implies p_{\hat{a}} = e^{\mu}_{\hat{a}} p_{\mu} = (\hat{E},p_{\hat{i}}).
\ee

We define spacetime gamma matrices $\Gamma^{\mu} = e_{\hat{a}}^{\mu} \Gamma^{\hat{a}}$ as a section of the bulk Clifford bundle determined by some fixed frame gamma matrices $\Gamma^{\hat{a}}$. These satisfy the Clifford algebra
\be
\label{cliffordalgebra}
\{ \Gamma^{\hat{a}}, \Gamma^{\hat{b}} \} = 2 \eta^{\hat{a}\hat{b}} \implies \{ \Gamma^{\mu}, \Gamma^{\nu} \} = 2 g^{\mu \nu}.
\ee
In this paper we will use the basis
\be
\label{gammaconventions}
\Gamma^{\hat{t}} = \begin{pmatrix} i & 0 \\ 0 & -i \end{pmatrix}, \ \ \Gamma^{\hat{i}} = \begin{pmatrix} 0 & \sigma^{\hat{i}} \\ \sigma^{\hat{i}} & 0 \end{pmatrix}.
\ee
We use hats on the Pauli matrices since they really belong to the tangent frame. Dirac conjugation is defined as usual, in terms of the spacetime gamma matrices: $\overline{\psi} = \psi^{\dagger} \Gamma^{t}$. 

We write derivatives transforming under both coordinate and gauge transformations,
\be
D_{\mu} = \partial_{\mu} - \frac{i}{4} \omega_{\mu,\hat{a}\hat{b}} \sigma^{\hat{a}\hat{b}} - i A_{\mu}.
\ee
The second term involves the spin connection, which has coefficients
\be
\omega_{\mu,\hat{a}\hat{b}} = e_{\nu,\hat{a}} e^{\lambda}_{\hat{b}} \Gamma^{\nu}_{\mu \lambda} - e_{\nu,\hat{a}} \partial_{\mu} e^{\nu}_{\hat{b}}
\ee
with $\Gamma^{\lambda}_{\mu \nu}$ the Christoffel symbols; the only non-vanishing components in the metric \eqref{bulkansatz} are
\begin{align}
\begin{split}
\omega_{t}^{\hat{t} \hat{z}} & = -\omega_{t}^{\hat{z} \hat{t}} = \frac{f'}{2 \sqrt{f g}}, \\
\omega_{x}^{\hat{z} \hat{x}} & = - \omega_{x}^{\hat{x} \hat{z}} = \omega_{y}^{\hat{z} \hat{y}} = - \omega_{y}^{\hat{y} \hat{z}} = \frac{1}{z^2 \sqrt{g}},
\end{split}
\end{align}
and the (tangent frame) spin operator is
\be
\sigma^{\hat{a}\hat{b}} = \frac{i}{2} \left[ \Gamma^{\hat{a}}, \Gamma^{\hat{b}} \right].
\ee
For the purpose of rescaling the Dirac spinor in section \ref{wkbsection}, note that an easy calculation gives the spin connection's contribution to the Dirac equation:
\be
\Gamma^{\mu} \omega_{\mu,\hat{a}\hat{b}} \sigma^{\hat{a}\hat{b}} = i \partial_z \ln [- (\det g) g^{zz} ] \Gamma^z.
\ee

\section{Fermions in magnetic fields in flat spacetime}

In this appendix we review some pertinent facts about the dynamics of relativistic fermions in a constant background magnetic field in flat spacetime. We first obtain the single-particle energy spectrum and eigenstates of the fermions. We use these to compute, in the grand canonical ensemble, the energy density, pressure, and charge density a gas of such particles. We obtain explicitly these thermodynamic quantities in the limit $T/\mu \ll B/\mu^2 \ll 1$, up to the leading order contributions in $B/\mu^2$. The zeroth order answer is that of a free degenerate Fermi gas while the first order corrections are periodic in $1/B$.

\subsection{Spectrum}

Consider a Dirac fermion $\psi$ with mass $m$ and charge $q$ propagating in four-dimensional Minkowski spacetime with metric $\eta_{\mu\nu} = (-1,1,1,1)$. The fermion is subject to a constant background magnetic field, which we assume to be in the $z$-direction, $F = B dx \wedge dy$. We will work in Landau gauge 
\be
\label{landaugauge}
A = -\mu dt + B x dy
\ee
with $\mu$ the chemical potential. The fermion dynamics is determined from the action
\begin{align}
{S}=\int d^4x \,i\bar\psi\left(\Gamma^{\mu} D_{\mu} -m\right)\psi, 
\end{align}
where $\Gamma^{\mu}$ can be taken in eg. the basis \eqref{gammaconventions}, and $D_\mu$ is the covariant derivative $D_\mu=\partial_\mu-i A_\mu$.

The energy spectrum of $\psi$ is determined by solving the Dirac equation
\begin{align}\label{flatdiraceq}
\left(\Gamma^{\mu} D_{\mu} -m\right)\psi=0.
\end{align}
The problem is time-independent and we can take $\psi(\mb{x},t) = e^{-i \omega t} u(\mb{x})$. Then we can rewrite the Dirac equation as a time-independent Schr\"{o}dinger problem,
\begin{align}
\begin{split}
H u & = E u \\
H & = - \Gamma^0 \left[ \Gamma^i (p_i - A_i) + i m \right],
\end{split}
\label{flatdiraceqsimple}
\end{align}
where as usual $p_i = - i \partial_i$, and we defined $E = \omega - \mu$. The spinor structure here is somewhat complicated, but notice that if we have a solution to the squared equation
\be
\label{diracsquare}
H^2 \chi = E^2 \chi
\ee
then the state
\be
\label{diracsolntrick}
u = (H+E) \chi
\ee
will solve the Dirac equation \eqref{flatdiraceqsimple}. In Landau gauge \eqref{landaugauge}, the square of the Hamiltonian is simply
\be
H^2 = m^2 + p_x^2 + (p_y - B x)^2 + p_z^2 - B \Sigma
\ee
where $\Sigma = -i \Gamma^x \Gamma^y = \sigma^z \otimes \mb{1}_{2 \times 2}$ is the spin operator. This is clearly block-diagonal and each $2 \times 2$ block is identical.

Solving \eqref{diracsquare} is simple. The problem is symmetric under translations in both $y$ and $z$, i.e. $p_y$ and $p_z$ commute with the Hamiltonian, so the eigenstates of $H^2$ are plane waves $\chi \sim e^{i p y + i q z}$. Define the ``magnetic algebra''
\begin{align}
\begin{split}
x & = \frac{1}{\sqrt{2B}} \left[ A + A^{\dagger} + \sqrt{\frac{2}{B}} p \right] \\
p_x & =- i \sqrt{\frac{B}{2}} \left[ A - A^{\dagger} \right].
\end{split}
\end{align}
These satisfy $[A,A^{\dagger}] = 1$ on functions of $x$. The constant shift corresponds to a shift of the $x$-origin by $p/B$. Clearly the eigenstates are of the form
\be
\chi = \chi_{pqn\pm}(\mb{x}) = N e^{i (p y + q z)} \psi_n(x-p/B) \zeta_{\pm}
\ee
where $\psi_n$ is the $n$th harmonic oscillator wavefunction, and $\zeta_{\pm}$ is a spin-$z$ eigenstate. This is the solution in either of the blocks, and we can take the other block to simply vanish. The energies satisfy
\be
E_{pqn\pm}^2 = m^2 + q^2 + 2 B \left(n + \frac{1}{2} \mp \frac{1}{2} \right).
\ee
The index $n$ is known as the Landau level of the state.

This spectrum has two notable features. First, the momentum along $y$ has dropped out, so the spectrum is continuously degenerate in that quantum number. More interestingly, the spin-dependent term has exactly the right magnitude so that
\be
E_{qn}^2 := m^2 + q^2 + 2Bn = E_{pqn+}^2 = E_{pq(n-1)-}^2,
\ee
i.e. the state at Landau level $n$ and spin up is degenerate with the state at level $n-1$ with spin down: the dipole potential of the cyclotron orbits exactly cancels that from the spin. Thus for fixed $p,q$, the energy spectrum is fourfold degenerate (counting the antiparticle states), except for the ground state $n=0, \sigma^z = +1$ which is only doubly-degenerate.

To write the explicit solutions to \eqref{flatdiraceq}, it is convenient to write
\be
H = \begin{pmatrix} m & \Oi \\ - \Oi & -m \end{pmatrix}
\ee
where
\be
\Oi = -\sqrt{\frac{B}{2}} \left[ (A - A^{\dagger}) \sigma^x - i (A + A^{\dagger}) \sigma^y \right] - i q \sigma^z.
\ee
This operator acts on the spin-oscillator part of the wavefunctions as
\begin{align}
\begin{split}
\Oi \psi_n \zeta_+ & = -\sqrt{2 B n} \psi_{n-1} \zeta_{-} - i q \psi_{n} \zeta_+  \\
\Oi \psi_n \zeta_- & = \sqrt{2 B (n+1)} \psi_{n+1} \zeta_{+} + i q \psi_{n} \zeta_-.
\end{split}
\end{align}

Now we can use this and the construction \eqref{diracsolntrick} to get the full solutions. For the particle states we can begin with the multiplet 
\be
\chi^1_{pqn} = \begin{pmatrix} \chi_{pqn+} \\ 0 \end{pmatrix}, \ \ \chi^2_{pqn} = \begin{pmatrix} \chi_{pq(n-1)-} \\ 0 \end{pmatrix}
\ee
both members with $E = E_{qn}$. Then, acting with $H + E$, we obtain the orthogonal pair of states
\begin{align}
\begin{split}
u^{1}_{pqn} & = \begin{pmatrix} (E_{qn} + m) \chi_{pqn+} \\ \sqrt{2 B n} \chi_{pq(n-1)-} + i q \chi_{pqn+} \end{pmatrix} \\
u^{2}_{pqn} & = \begin{pmatrix} (E_{qn} + m) \chi_{pq(n-1)-} \\ -\sqrt{2 B n} \chi_{pqn+}  - i q \chi_{pq(n-1)-}  \end{pmatrix}.
\end{split}
\end{align}
One can obtain quite similar formulae for a multiplet of antiparticle states, starting with the $\chi$'s in the lower block rather than the upper block.

From here out we will sometimes use the schematic index $\alpha = \{ p q n i \}$ to label these states and energies. In order to normalize the second-quantized Hamiltonian $H = \sum_{\alpha} E_{\alpha} a_{\alpha}^{\dagger}a_{\alpha}$, we want to normalize $\int d^3x u_{\alpha}^{\dagger} u_{\beta} = \delta_{\alpha \beta}$. For simplicity let us work in a box of volume $V = L^3$.\footnote{Although we still follow the common practice of taking $\int dx \psi^*_n(x) \psi_m(x) = \delta_{nm}$ for the oscillator wavefunctions.}  This sets
\be
|N|^2  = \left[ 2 L^2 E_{qn}(E_{qn} + m) \right]^{-1}.
\ee

\subsection{$T=0$ statistical mechanics}
Consider now a large number of these fermions. Let's study the system in the grand canonical ensemble where the chemical potential, magnetic field and temperature are kept fixed. 

The energy spectrum 
\be
E_{qn}^2 = m^2 + q^2 + 2Bn
\ee
has a huge degeneracy, as remarked above. To regulate the computations, note that the totally degenerate momentum $p$ along the $y$-axis is bounded from above, $|p| \leq L B/2$, due to the fact that the shifted center of the $x$-coordinate harmonic oscillator $x_0=p/B$ must stay inside the box.\footnote{Another way to see this is that in axial gauge $A \sim B (x dy - y dx)$ the wavefunctions are localized annuli on the plane, which must remain sufficiently within the box.} The states with $n \geq 1$ are all doubly degenerate due to spin. So, at each level $n$, the total degeneracy becomes
\begin{align}
\label{degeneracy}
g_n = 2 \pi A d_n, \ \ d_n = \frac{B}{(2\pi)^2} (2-\delta_{n,0})
\end{align}
where we took the usual $L/2\pi$ normalization for the plane wave factor, and the reason for defining $d_n$ this way will become clear shortly.

Let us fix some notation by considering the physics at strictly zero temperature. For $T=0$, all states with $E_{qn} = \omega_{qn} - \mu < 0$ will be occupied, and none of the others. In each level $n$ we define an effective mass
\begin{align}
m_n^2 = m^2 + 2 B n.
\end{align}
For fixed $\mu > 0$ we have that, in a given Landau level $n$, the momentum states that are occupied are those with
\be
q^2 \leq q^2_n := \mu^2 - m_n^2.
\ee 
Similarly, there will be a maximum Landau level with occupied states, given by the floor of the continuous quantity $\nf$. Let $\Delta m^2 = \mu^2 - m^2$, then let
\be
\label{nfdef}
\nf = \frac{\Delta m^2}{2B}.
\ee
If the highest Landau level only has its $q=0$ states occupied, then $\nf$ will be an integer equal to that level. As $B$ or $\mu$ vary, $\nf$ varies continuously and tracks the fraction of the uppermost Landau level that is filled. Shortly we will see that, in the limit where $T/\mu \lesssim B/\mu^2 \ll 1$, the thermodynamic quantities are oscillatory in $\nf \sim 1/B$.

In the strict $T=0$ limit, one can explicitly sum over the occupied states $\omega \leq \mu$ to obtain the usual thermodynamic quantities. The cleanest way to do it is to compute the expectation of the stress energy tensor components in the usual Fermi sea state. Expand the Dirac field
\begin{align}
\begin{split}
\psi(x) & = \sum_{\alpha} e^{-i  \omega_{\alpha} t} u_{\alpha}(\mathbf{x}) a_{\alpha} + e^{i  \omega_{\alpha} t}v_{\alpha}(\mathbf{x}) b_{\alpha}^{\dagger},\\
\psi^{\dagger}(x) & =  \sum_{\alpha} e^{-i  \omega_{\alpha} t} v^{\dagger}_{\alpha}(\mathbf{x}) b_{\alpha} + e^{i  \omega_{\alpha} t} u^{\dagger}_{\alpha}(\mathbf{x}) a^{\dagger}_{\alpha}.
\end{split}
\label{modeexpansionflat}
\end{align}
Fix a chemical potential $\mu > 0$. The Fermi sea is the state
\be
\ket{\mu} = \prod_{0 \leq \omega_{\alpha} \leq \mu} a_{\alpha}^{\dagger} \ket{0},
\ee
where the product is ordered with energy increasing from right to left. Due to Fermi statistics, this state has the algebraic properties
\be
\label{fermisea}
\begin{split}
\braket{ \mu | a^{\dagger}_{\alpha} a_{\alpha'} | \mu } & = \delta_{\alpha,\alpha'} \Theta(\omega_{\alpha} \leq \mu) \\
\braket{ \mu | a_{\alpha} a^{\dagger}_{\alpha'} | \mu } & = \delta_{\alpha,\alpha'} \left[ 1 - \Theta(\omega_{\alpha} \leq \mu) \right].
\end{split}
\ee
When we take vacuum expecation values of bilinear operators like the current and stress tensor, we subtract the contribution from the usual vacuum state $a_{\alpha} \ket{ 0 } = 0$, and we drop any anti-particle contributions, yielding finite answers.

We are now in a position to compute the expectation values of the current and energy-momentum tensor operators. Let us start with the expectation value of the current. Using \eqref{modeexpansionflat}, \eqref{fermisea}, one easily obtains, for example, the charge density
\begin{align}
\sigma = \langle J^{t}\rangle = \langle \mu |  \overline{\psi} \Gamma^{t} \psi | \mu \rangle = \sum_{\rm sea} u_{\alpha}^{\dagger} u_{\alpha}.
\label{currenttimecomponent}
\end{align}
Here the sum over the sea means a sum over frequencies $0 \leq \omega_{\alpha} \leq \mu$, i.e.
\be
\sum_{sea} = \frac{B V}{(2\pi)^2} \sum_{i=1,2} \sum_{n=0}^{n_F} \int_{-q_n}^{q_n} dq,
\ee
with $n_F, q_n$ defined above, and where we have taken the continuum limit for the $z$-axis momenta. Using the explicit states $u_{\alpha}$ given above, and using the fact that our state is translation invariant in every direction to write $\int d^3x/V = 1$, this reduces to
\be
\sigma = 2 \sum_{n=0}^{n_F} d_n q_n.
\ee 
Going through quite similar manipulations, one easily finds that $\braket{ J^{i}} = 0$ along the spatial axes. 

The expectation values of the stress tensor components are a bit more computationally involved but straightforward. One obtains
\be
\braket{ T_{\mu \nu}^F } = \begin{pmatrix} \rho & & & \\ & p_{\perp} & & \\ & & p_{\perp} & \\ & & & p_{z} \end{pmatrix}
\ee
where these thermodynamic functions are given by
\begin{align}
\begin{split}
\rho & = \sum_{n=0}^{n_F} d_n \left[ \mu q_n + m_n^2 \ln \left( \frac{q_n + \mu}{m_n} \right) \right] \\
p_z & = \sum_{n=0}^{n_F} d_n \left[ \mu q_n - m_n^2 \ln \left( \frac{q_n + \mu}{m_n} \right) \right] \\
p_{\perp} & = 2 \sum_{n=0}^{n_F} d_n \left[ 2 B n  \ln \left( \frac{q_n + \mu}{m_n} \right) \right].
\end{split}
\label{thermoexact}
\end{align}
The sum over Landau levels is not possible to evaluate in closed form. However, one can work at small finite temperaute, and perform some approximations in the exact partition function which give tractable and interesting formulas. We now turn to this.

\subsection{Equation of state, de Haas-van Alphen effect}
\label{dhvaappendix}

In the absence of magnetic field, the energy spectrum is given by $\omega^2=m^2+\vec k^2$, and states whose energies are below the Fermi energy $\omega_{\rm F}^2=\mu^2=m^2+\vec k_{\rm F}^2$ are all occupied. This implies that the $B\to 0$ limit should be taken provided that $2n_{\rm F} B = \mu^2-m^2$ remains fixed. In this limit, the number of Landau levels with occupied states below the Fermi surface becomes infinite, so we can approximate the sum over Landau levels by an integral. 

In the limit that $T/\mu \leq B/\mu^2 \ll 1$, thermodynamic quantities develop a part which is periodic as a function of $1/B$. This is the de Haas-van Alphen effect. One can see it by thinking of the fermion states as cyclotron orbits on the plane, whose radii are quantized in units of the flux. The total thermodynamics is then the sum of a degenerate Fermi gas plus corrections that oscillate with the external magnetic field, as well as terms monotonic in the magnetic field. These oscillations can be easily seen by simply plotting the exact thermodynamics \eqref{thermoexact} as a function of $B$. In what follows we get an analytical approximation for this effect.

Working at arbtirary temperature and field, we can write down the exact grand potential. Taking the continuum limit of the $z$-axis momentum $q$, the grand potential takes the form
\begin{align}
\label{omegaexact}
\Omega & = - T \ln Z \\
& = -V T  \sum_{n=0}^{\infty} d_n \int_{-\infty}^{\infty} dq \ln \left[ 1+e^{-(\omega - \mu)/T}\right].
\end{align}
In order to analytically obtain the oscillatory behavior of the thermodynamic quantities, we can use Poisson's summation formula\footnote{This comes from writing the Dirac comb as its Fourier series, $\sum_{n=-\infty}^{\infty} \delta(x-n) = \sum_{k=-\infty}^{\infty} e^{2\pi i k x},$ and integrating against a test function $f$ on $x \in [0, \infty]$. The Poisson summation identity is then
\begin{align}
\label{poisson}
\frac{1}{2} \sum_{n=0}^{\infty} (2-\delta_{n,0}) f(n)&= \int_{0}^{\infty} dx \ f(x) 
+ 2\,\text{Re} \sum_{k=1}^{\infty} \int_{0}^{\infty} dx \ e^{2\pi i k x} f(x).
\end{align}} 
to write the grand potential $\Omega$ in terms of oscillating functions. Let $f(n) = \int dq \ln ...$. Now apply Poisson summation to this. 

The first term is non-oscillatory, independent of $B$, and gives the grand potential of a free Fermi gas
\be
\Omega_0 = -\frac{\Delta m^2 V}{8 \pi^2} \left[ \frac{2}{3} \mu \Delta m - \frac{1}{3} \Delta m^2 \right].
\ee
This leads to the zeroth-order contribution to $\sigma$, $\rho$, $p_z$ and $p_\perp$:
\begin{align}
\begin{split}
\sigma_0 &=\frac{\Delta m^3}{3\pi^2} \\
\rho_0 & = \frac{1}{8\pi^2}\left[\mu(2\mu^2-m^2) \Delta m - m^4\ln{ \left(\frac{\mu + \Delta m}{m}\right)} \right] \\
p_0&=\frac{1}{24\pi^2} \left[\mu(2\mu^2-5m^2) \Delta m +3 \ln{\left(\frac{\mu + \Delta m}{m} \right)}  \right].
\end{split}
\end{align}
Since the equation of state at this order is independent of $B$, the pressure $p_0$ is isotropic.

The oscillatory integrals contain all the information about the magnetic field. The integrals produce both oscillatory and monotonic functions of $B$, but the latter are supressed by an extra factor $B/\mu^2$, so at small $B/\mu^2$ we get the dominant oscillatory behavior
\begin{align}
\begin{split}
\Omega_{\rm osc} & = \frac{VT B^{3/2}}{2\pi^2} \sum_{k=1}^{\infty}\frac{\cos(2\pi k\nf -\pi/4)}{k^{3/2} \sinh(2\pi^2k \mu T/B)} \\
& \approx  \frac{V B^{5/2}}{4\pi^4\mu} \sum_{k=1}^{\infty}\frac{1}{k^{5/2} }\cos\Big(2\pi k\nf -\frac{\pi}{4}\Big).
\end{split}
\end{align}
where $\nf$ has been defined in \eqref{nfdef}, and in the second line we expanded around low temperatures $T/\mu \ll 1$. 

Thus, at zero temperature, and to leading order in the magnetic field, one obtains
\begin{align}
\begin{split}
\sigma\,&=\sigma_0+\frac{B^{3/2}}{\pi^3} \sum_{k=1}^{\infty}\frac{1}{k^{3/2} }\sin\Big(2\pi k\nf -\frac{\pi}{4}\Big) \\
\rho\,&=\rho_0+\frac{\mu B^{3/2}}{\pi^3} \sum_{k=1}^{\infty}\frac{1}{k^{3/2} }\sin\Big(2\pi k\nf -\frac{\pi}{4}\Big) \\
p_z&=p_{0}-\frac{B^{5/2}}{4\pi^4\mu} \sum_{k=1}^{\infty}\frac{1}{k^{5/2} }\cos\Big(2\pi k\nf -\frac{\pi}{4}\Big) \\
p_\perp&=p_{0}+\frac{B^{3/2}}{2\pi^3\mu} \Delta m^2 \sum_{k=1}^{\infty}\frac{1}{k^{5/2} }\sin\Big(2\pi k\nf -\frac{\pi}{4}\Big).
\end{split}
\label{flatthermo}
\end{align}

\pagebreak

\bibliography{dyonstar-new}

\end{document}